\documentclass[english,aps,preprint]{revtex4}
\usepackage[T1]{fontenc}
\usepackage[latin9]{inputenc}
\usepackage{amsmath}
\usepackage{amssymb}
\usepackage{graphicx}

\usepackage{CJK}

\makeatletter
\@ifundefined{textcolor}{}
{%
 \definecolor{BLACK}{gray}{0}
 \definecolor{WHITE}{gray}{1}
 \definecolor{RED}{rgb}{1,0,0}
 \definecolor{GREEN}{rgb}{0,1,0}
 \definecolor{BLUE}{rgb}{0,0,1}
 \definecolor{CYAN}{cmyk}{1,0,0,0}
 \definecolor{MAGENTA}{cmyk}{0,1,0,0}
 \definecolor{YELLOW}{cmyk}{0,0,1,0}
}

\usepackage{babel}

\makeatother

\usepackage{babel}
\begin{document}
\begin{CJK*}{GB}{}

\title{Path integral molecular dynamics for anyons, bosons and fermions}

\author{Yunuo Xiong}
\affiliation{College of Science, Zhejiang University of Technology, Hangzhou 31023, China}

\author{Hongwei Xiong}
\email{xionghw@zjut.edu.cn}
\affiliation{College of Science, Zhejiang University of Technology, Hangzhou 31023, China}

\begin{abstract}

In this article we develop a general method to numerically calculate physical properties for a system of anyons with path integral molecular dynamics. We provide a unified method to calculate the thermodynamics of identical bosons, fermions and anyons. Our method is tested and applied to systems of anyons, bosons and fermions in a two-dimensional harmonic trap. We also consider a method to calculate the energy for fermions as an application of the  path integral molecular dynamics to simulate the anyon model.

Keywords: anyons; bosons; fermions; path integral; molecular dynamics
\end{abstract}
\maketitle
\end{CJK*}

\section{introduction}

Quantum many-body systems are fascinating and reveal much of the interesting phenomenon of nature. In particular, systems consisting of identical particles are of great interests since quantum mechanics dictates that fundamental exchange symmetries be present in such systems. Insightful effects such as Berezinskii-Kosterlitz-Thouless transition \cite{Kosterlitz,Hadzibabic} are found resulting from the exchange symmetry between identical particles. Traditionally, people have considered only two kinds of fundamental exchange symmetry in a system of identical particles, with the wave function either be completely symmetric or anti-symmetric, corresponding to bosons and fermions respectively. However, when studying quantum many-body systems in two dimensions, researchers have found that other kinds of exchange symmetry are possible \cite{Leinaas}. This leads to the discovery of anyons \cite{Leinaas,FluxW,a-wilczek,Wilczek,khare}. Anyons manifest novel phenomenon such as fractional quantum Hall effects \cite{Tsui} and their existences in two dimensional systems are shown by recent experiments \cite{Bar, Nak,Nak2}.
\par
Aside from experimental study of many-body quantum systems, numerical simulations of such systems are also desirable. There are two main ways to perform the simulations based on path integral formalism \cite{feynman,kleinert}: path integral Monte Carlo \cite{CeperRMP,boninsegni1,boninsegni2,Dornheim} and path integral molecular dynamics (PIMD) \cite{chandler,Parrinello,Tuckerman}. Path integral Monte Carlo has been applied successfully to large bosonic systems \cite{boninsegni1,boninsegni2}, by sampling directly the partition function corresponding to the system under study. However, PIMD has found few applications in the field of identical particles until recently, when an efficient polynomial algorithm \cite{Hirshberg} to perform PIMD has been developed, enabling applications of PIMD to large bosonic system \cite{Hirshberg,Xiong,XiongSpinor} and fermionic system \cite{HirshbergFermi,Xiong1}. Both of these approaches suffer from severe efficiency issues when applied to identical fermions, due to the so-called fermion sign problem \cite{ceperley,loh,troyer,lyubartsev,vozn,Yao1,Yao2,Yao3}.
\par

In this paper we extend PIMD to perform numercial simulations for a system of identical anyons. In particular, we calculate the energies and densities for anyons and study their dependences on the statistical properties for the anyons. Our work provided a unified method to calculate various properties of anyons, bosons and fermions.
The paper is organized as follows: In Sec. \ref{theory}, we start by reviewing path integral formulations for anyons. We then present PIMD and show how to extend the technique to perform simulations for a system of identical anyons. In Sec. \ref{result}, we present our simulation results for the energies and densities for anyons and study their dependences on the statistical properties of the anyons. 
In Sec. \ref{alleviate}, by taking advantage of the extra freedom in choosing the statistical properties for anyons, we suggest a way to calculate the energy of a few fermions by an extrapolation with a simple function based on reasonable physical considerations.
In Sec. \ref{summary}, we give a concise summary and discussion of our work.

\section{theory}
\label{theory}

\subsection{Path integral for anyons, bosons and fermions}

In this paper, we will use "boson" gauge to consider the anyons in a general way \cite{Myrheim}. The derivation from Bose-Einstein statistics in this bosonic gauge is due to a "statistics" field, which is a vector potential based on the flux-tube model for anyons by Wilczek \cite{FluxW}. The Bose statistics may change continuously into Fermi statistics when the statistics angle $\theta$ changes from boson value $0$ to fermion value $\pi$. The fractional statistics is the situation of $0<\theta<\pi$. In this case, it is anticipated that there would be a general numerical method which can be applied to both bosons and fermions in a unified way.

\begin{figure}
\centering \includegraphics[width=0.8\textwidth]{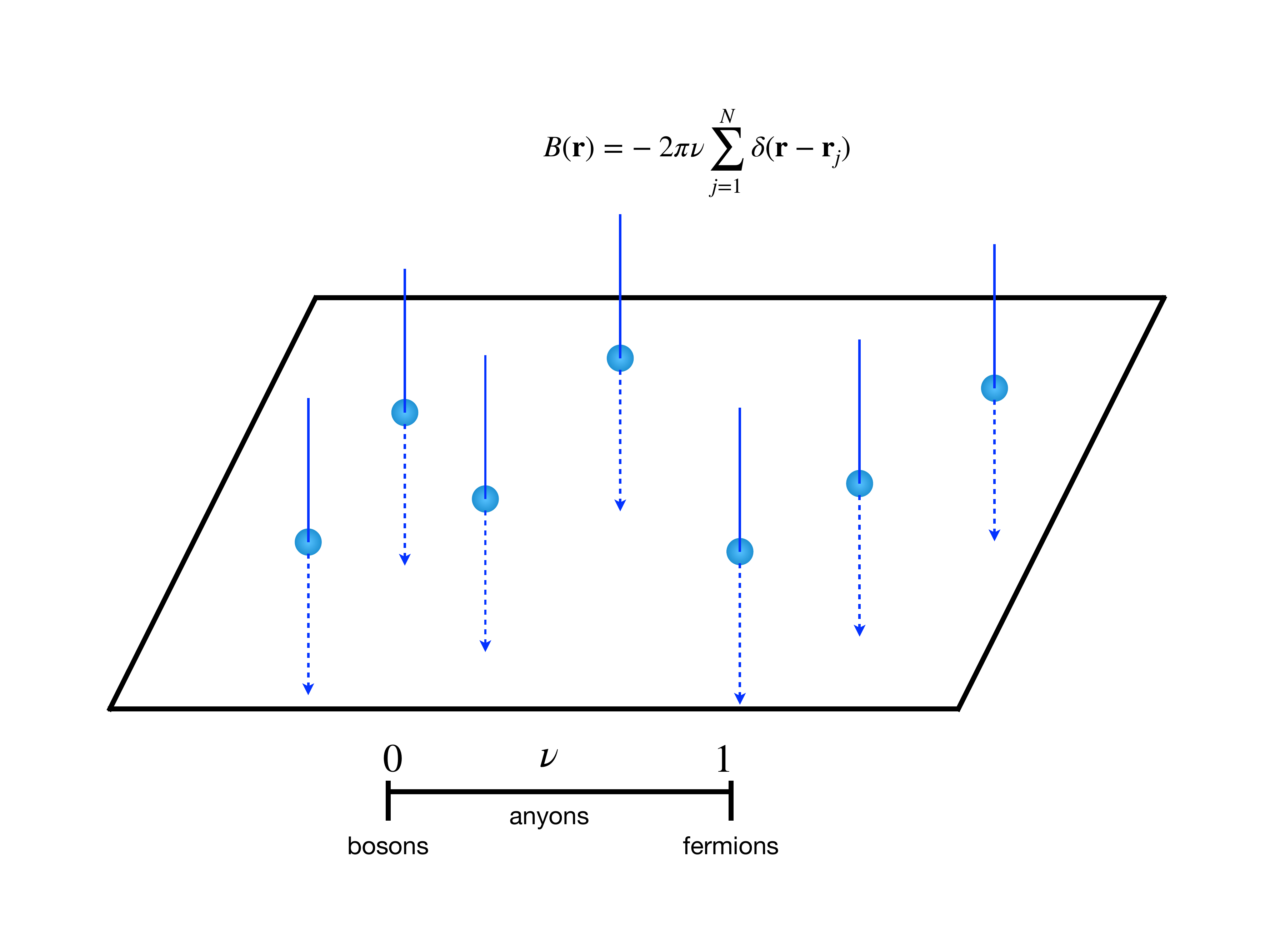} 
\caption{The flux-tube model for anyons.}
\label{anyon1}
\end{figure}

In the flux-tuble model for anyons \cite{FluxW}, the anyons turn out to be point particles having a fictitious magnetic flux.   
In this case, the flux is located exactly at the position of the $j$th particle, and the fictitious magnetic field for this particle is
\begin{equation}
B(\mathbf r)=-2\pi \nu \delta (\mathbf r-\mathbf r_j).
\end{equation}
Here $\nu$ is a parameter for the statistical properties of the anyon. $\nu=0$ for bosons and $\nu=1$ for fermions. The statistics angle $\theta=\nu\pi$. For $N$ anyons, the total magnetic field is 
\begin{equation}
B(\mathbf r)=-2\pi \nu \sum_{j=1}^N \delta (\mathbf r-\mathbf r_j).
\end{equation}

From the relation between $B$ and the vector potential $\mathbf A$
\begin{equation}
B=\partial_x A_y-\partial_y A_x,
\end{equation}
we have the following vector potential at $\mathbf r$ generated by the $j$th particle,
\begin{equation}
A_x^j(\mathbf r)=\nu\frac{y-y_j}{(x-x_j)^2+(y-y_j)^2},\nonumber
\end{equation}
\begin{equation}
A_y^j(\mathbf r)=-\nu\frac{x-x_j}{(x-x_j)^2+(y-y_j)^2}.
\end{equation}

The N-particle Hamiltonian is
\begin{equation}
\hat H_N=\frac{1}{2m} \sum_{j=1}^N \left(\hat {\mathbf p}_j- \hbar{\mathbf a}_j \right)^2+V.
\end{equation}
Here $\hat{\mathbf p}_j$ is the canonical momentum of the $j$th particle. $V$ is the interaction potential including both the external potential and interaction potential between anyons. The statistical interaction is due to the vector potential ${\mathbf a}_j $ for the $j$th particle. Because this vector potential is due to all other particles, we have
\begin{equation}
a_{jx}(\mathbf r_j)=\nu\sum_{k\neq j}\frac{y_j-y_k}{(x_j-x_k)^2+(y_j-y_k)^2},\nonumber
\end{equation}
\begin{equation}
a_{jy}(\mathbf r_j)=-\nu\sum_{k\neq j}\frac{x_j-x_k}{(x_j-x_k)^2+(y_j-y_k)^2}.
\end{equation}

\subsection{The path integral for anyons}
The partition function for $N$ anyons is
\begin{equation}
Z_A(\beta)=Tr e^{-\beta\hat H_N}.
\end{equation}
Here $\beta=1/k_BT$ with $k_B$ the Boltzmann constant and $T$ the temperature of the system.
Because we use the bosonic gauge to consider the anyons, we have
\begin{equation}
Z_A(\beta)=\frac{1}{N!}\sum_{p\in S_N} \int d\mathbf r_1\cdots d\mathbf r_N \left<p\{\mathbf r_1,\cdots,\mathbf r_N\}\right|e^{-\beta\hat H_N} \left|\mathbf r_1,\cdots,\mathbf r_N\right>.
\end{equation}
Here $p$ denotes $N!$ permutation of $S_N$ permutation group. The integral is about the usual $\mathbf R^{dN}$ space.

We divide $\beta$ into $P$ intervals (let us denote the interval length by $\Delta\beta=\beta/P$) and expand each of the evolution operators into products of the form
\begin{equation}
e^{-\beta \hat H}=e^{-\Delta\beta \hat H}\cdots e^{-\Delta\beta \hat H}.
\end{equation}
In the limit of $P\rightarrow \infty$, we have
\begin{equation}
Z_A(\beta)=\frac{1}{N!}\sum_{p\in S_N}\int_{\mathcal C(p)} \mathcal D[\mathbf r_1 (\tau),\cdots,\mathbf r_N(\tau)]e^{-\frac{S}{\hbar}-i\theta Q}.
\label{Apartition}
\end{equation}
Here 
\begin{equation}
S=\int_0^{\hbar\beta} d\tau\left(\frac{m}{2}\sum_{j=1}^N\left|\frac{d\mathbf r_j}{d\tau}\right|^2+V(\mathbf r_1\cdots \mathbf r_N)\right),
\end{equation}
and
\begin{equation}
-\theta Q=-\pi \nu Q=\sum_{j=1}^N\int d\mathbf r_j\cdot \mathbf a_j(\mathbf r_j). 
\label{Qintegral}
\end{equation}
This equation stands for the sum of a series of line integrals. To evaluate such line integrals some parametrization of the curve must be chosen. Of course the variable $\tau$ provides a natural parametrization and this was used in our numerical calculation, but generally any parametrization works and leads to the same result for the integral. We chose to write the line integral in the general form in Eq. (\ref{Qintegral}), whereas the discrete form based on $\tau$ parametrization will be presented in Eq. (\ref{dis}).
In the above equation for $Z_A(\beta)$, the domain of integration $\mathcal C(p)$ represents $\{\mathbf r_1(\hbar\beta),\cdots \mathbf r_N(\hbar\beta)\}=p\{\mathbf r_1(0),\cdots,\mathbf r_N(0)\}$.

In our numerical calculations, $P$ is always finite. In this case, the approximate  $Z_A(\beta)$ can be written as
\begin{equation}
Z_A(\beta)=\left(\frac{mP}{2\pi\hbar^2\beta}\right)^{PdN/2} \int e^{-\beta(V_B^{(N)}+\frac{1}{P}U)}e^{-i \theta \Phi}d\mathbf R_1...d\mathbf R_N,
\label{partition}
\end{equation}
where $\mathbf R_i$ represents the collection of ring polymer coordinates $(\mathbf r_i^1,...,\mathbf r_i^P)$ corresponding to the $i$th particle. The system considered has $d$ spatial dimensions. For anyons considered in this work, $d=2$. $V_B^{(N)}$ considers exchange effects by describing all the possible ring polymer configurations. $U$ is the interaction between different particles, which is given by
\begin{equation}
U = \sum_{l=1}^P V(\mathbf r_1^l,...,\mathbf r_N^l).
\end{equation}
In addition,
\begin{equation}
-\theta\Phi(\mathbf R_1,...,\mathbf R_N)=\sum_{j=1}^N\sum_{l=1}^{P-1}(\mathbf r_j^{l+1}-\mathbf r_j^l)\cdot\mathbf a_j(\mathbf r_1^l,...,\mathbf r_N^l),
\label{dis}
\end{equation}
where $\mathbf a_j(\mathbf r_1^l,...,\mathbf r_N^l)$ is given by
\begin{equation}
a_{jx}(\mathbf r_1^l,...,\mathbf r_N^l)=\nu\sum_{k\neq j}\frac{r_{jy}^l-r_{ky}^l}{(r_{jx}^l-r_{kx}^l)^2+(r_{jy}^l-r_{ky}^l)^2},\nonumber
\end{equation}
\begin{equation}
a_{jy}(\mathbf r_1^l,...,\mathbf r_N^l)=-\nu\sum_{k\neq j}\frac{r_{jx}^l-r_{kx}^l}{(r_{jx}^l-r_{kx}^l)^2+(r_{jy}^l-r_{ky}^l)^2}.
\end{equation}

The expression of $V_B^{(N)}$ is the key to consider the exchange effects of identical bosons, which will be introduced in due course.

\subsection{Path integral molecular dynamics}
To carry out the numerical calculations, we should give the expression of $V_B^{(N)}$. Fortunately, in Ref. \cite{Hirshberg}, it is given for path integral molecular dynamics of bosons. In our bosonic gauge for anyons, because the factor $\theta Q$ is symmetric about the exchange of $\mathbf r_i$ and $\mathbf r_j$, the recursion formula of $V_B^{(N)}$ can be used without any change.

The recursion formula is
\begin{equation}
e^{-\beta V_B^{(\alpha)}}=\frac{1}{\alpha}\sum_{k=1}^\alpha e^{-\beta(E_\alpha^{(k)}+V_B^{(\alpha-k)})},
\label{recursion}
\end{equation}
where $V_B^{(0)}=0$ and $E_\alpha^{(k)}$ is given by
\begin{equation}
E_\alpha^{(k)}(\mathbf R_{\alpha-k+1},...,\mathbf R_\alpha)=\frac{1}{2}m\omega_P^2\sum_{l=\alpha-k+1}^\alpha\sum_{j=1}^P(\mathbf r_l^{j+1}-\mathbf r_l^j)^2,
\end{equation}
where $\mathbf r_l^{P+1}=\mathbf r_{\alpha-k+1}^1$ if $l=\alpha$ and $\mathbf r_l^{P+1}=\mathbf r_{l+1}^1$ otherwise. $E_\alpha^{(k)}$ may be interpreted as partial configuration of particles in which the last $k$ particles are fully connected. In particular, $E_N^{(N)}$ represents the longest ring configuration for $N$ particles. 

From the above recursion formula, we may get
\begin{equation}
V_B^{(1)}\rightarrow V_B^{(2)}\rightarrow \cdots  V_B^{(\alpha)}\cdots \rightarrow V_B^{(N)}.
\end{equation}

\par
Therefore we see that the recursive formula allows us to compute contributions from all ring polymer configurations correctly. Moreover, it is easy to see that evaluating $V_B^{(N)}$ only takes $O(PN^3)$ time, an exponential speedup over simply summing all configurations. 

We will use molecular dynamics and the massive and separate Nos\'e-Hoover chains \cite{Nose1,Nose2,Hoover,Martyna,Berne,Jang} to get the distribution $e^{-\beta(V_B^{(N)}+\frac{1}{P}U)}$.
To perform molecular dynamics we need the gradient of $V_B^{(N)}$, which may also be computed recursively from Eq. (\ref{recursion}) as
\begin{equation}
-\nabla_{\mathbf r_l^j}V_B^{(\alpha)}=-\frac{\sum_{k=1}^\alpha\left[\nabla_{\mathbf r_l^j}E_\alpha^{(k)}+\nabla_{\mathbf r_l^j}V_B^{(\alpha-k)}\right]e^{-\beta(E_\alpha^{(k)}+V_B^{(\alpha-k)})}}{\sum_{k=1}^\alpha e^{-\beta(E_\alpha^{(k)}+V_B^{(\alpha-k)})}},
\end{equation}
where $\nabla_{\mathbf r_l^j}E_\alpha^{(k)}$ is
\begin{equation}
\nabla_{\mathbf r_l^j}E_\alpha^{(k)}=m\omega_P^2\left[(\mathbf r_l^j-\mathbf r_l^{j+1})+(\mathbf r_l^j-\mathbf r_l^{j-1})\right],
\end{equation}
with the boundary conditions $\mathbf r_l^{P+1}=\mathbf r_{\alpha-k+1}^1$ if $l=\alpha$ and $\mathbf r_l^{P+1}=\mathbf r_{l+1}^1$ otherwise; $\mathbf r_l^{0}=\mathbf r_{\alpha}^P$ if $l=\alpha-k+1$ and $\mathbf r_l^{0}=\mathbf r_{l-1}^P$ otherwise. If $l$ is outside the interval $[\alpha-k+1,\alpha]$ then $\nabla_{\mathbf r_l^j}E_\alpha^{(k)}=0$. By recursion calculation from $\alpha=1$ to $\alpha=N$, we may get $-\nabla_{\mathbf r_l^j}V_B^{(N)}$ for the numerical calculations of the molecular dynamics.
\par
The above equations completely define a molecular dynamics algorithm for particles which scales as $O(PN^3)$, enabling the application of PIMD to large particle system. In order to extract physical quantities from our simulations we use various estimators. For example, from the following equation
\begin{equation}
\left<E\right>=-\frac{1}{Z_A{(\beta)}}\frac{\partial Z_A{(\beta)}}{\partial\beta},
\end{equation}
the average energy is given by
\begin{equation}
\left<E\right>=\frac{PdN}{2\beta}+\frac{\left<U\right>}{P}+\left<V_B^{(N)}+\beta\frac{\partial V_B^{(N)}}{\partial\beta}\right>.
\end{equation}
$V_B^{(N)}+\beta\frac{\partial V_B^{(N)}}{\partial\beta}$ may be evaluated as
\begin{equation}
V_B^{(N)}+\beta\frac{\partial V_B^{(N)}}{\partial\beta}=\frac{\sum_{k=1}^N[V_B^{(N-k)}+\beta\frac{\partial V_B^{(N-k)}}{\partial\beta}-E_N^{(k)}]e^{-\beta(E_N^{(k)}+V_B^{(N-k)})}}{\sum_{k=1}^Ne^{-\beta(E_N^{(k)}+V_B^{(N-k)})}},
\end{equation}
with $V_B^{(0)}+\beta\frac{\partial V_B^{(0)}}{\partial\beta}=0$.
\par
The average density is simply given by
\begin{equation}
\rho(\textbf{r})=\left<\frac{1}{P}\sum_{j=1}^P\sum_{k=1}^N\delta(\mathbf r_l^j-\textbf{r})\right>.
\end{equation}

It is worthy to point out that the statistical interaction term $e^{-i\theta Q}$ will not contribute to the expression of the energy estimator and density estimator. The key reason why $e^{-i\theta Q}$ does not affect the energy estimator is that the magnetic field in the region beyond the point where all particles are located is $0$.

Nevertheless,  $e^{-i\theta Q}$ will influence the average energy and average density. In our molecular dynamics simulation, we only get the distribution of $e^{-\beta(V_B^{(N)}+\frac{1}{P}U)}$. In this case, assume that $<>'$ denotes the average value in terms of this distribution, for an observable $\hat A$, we have
\begin{equation}
\left<\hat A\right>=\frac{\left<\epsilon_A e^{-i\theta Q}\right>'}{\left<e^{-i\theta Q}\right>'}.
\end{equation}
Here $\epsilon_A $ represents the estimator for $\hat A$.

Because the N-particle Hamiltonian operator is hermitian, for another hermitian operator $\hat A$, the average value
$\left<\hat A\right>$ should be real. This provides the general criteria to judge whether the numerical result is reasonable. Of course, for $\theta\neq 0$, we will have sign problem. Because $\theta=\pi$ is the situation of fermions, this means the famous fermion sign problem \cite{ceperley,loh,troyer,lyubartsev,vozn,Yao1,Yao2,Yao3}. Because $\theta=0$ is the situation of bosons, this means that bosons will not experience sign problem. The sign problem will become more severe with the increasing of $\theta$ between $0$ and $\pi$. Compared with the result of Ref. \cite{Hirshberg} for bosons and Ref. \cite{HirshbergFermi} for fermions, here we give a general formula for anyons, and provide the chance to study the continuous transition from bosons to fermions.

\section{Thermodynamics and sign problem in PIMD for anyon model}
\label{result}

In all our importance sampling, we use massive
Nos\'e-Hoover chain \cite{Nose1,Nose2,Hoover,Martyna,Berne,Jang} to establish constant temperature for our simulations,
where each separate degree of freedom of our system has been coupled
to a Nos\'e-Hoover thermostat. Specifically, we use the number of beads $P=12$ and $2\times10^{7}$
MD steps to achieve convergence. In all of the following we checked convergence with respect to the number of beads and MD steps performed. For details of how to assure the convergence and the general strategy to choose the number of beads and MD steps, one may refer to the supplementary material in \cite{Hirshberg} and also our previous works \cite{Xiong, Xiong1}.

Without the loss of generality, we consider the following interaction potential
\begin{equation}
V(\mathbf r_1,\cdots,\mathbf r_N)=\frac{1}{2}m\omega^2\sum_{j=1}^N |\mathbf r_j|^2+\sum_{l>k}^N\frac{\lambda}{|\mathbf r_l-\mathbf r_k|}.
\end{equation}
We will use the length unit $l_0=\sqrt{\hbar/m\omega}$ and energy unit $E_0=\hbar\omega$. In this case, the dimensionless $\tilde \beta$ is
\begin{equation}
\tilde \beta=\beta E_0,
\end{equation}
while the dimensionless $\tilde\lambda$ is
\begin{equation}
\tilde\lambda=\frac{\lambda}{E_0l_0}.
\end{equation}
We have
\begin{equation}
\frac{V}{E_0}=\frac{1}{2}\sum_{j=1}^N |\tilde{\mathbf r}_j|^2+\sum_{l>k}^N \frac{\tilde\lambda}{|\tilde{\mathbf r}_l-\tilde{\mathbf r}_k|},
\end{equation}
with $\tilde{\mathbf r}_k=\mathbf r_k/l_0$.

\begin{figure}[htbp]
\begin{center}
 \includegraphics[width=0.75\textwidth]{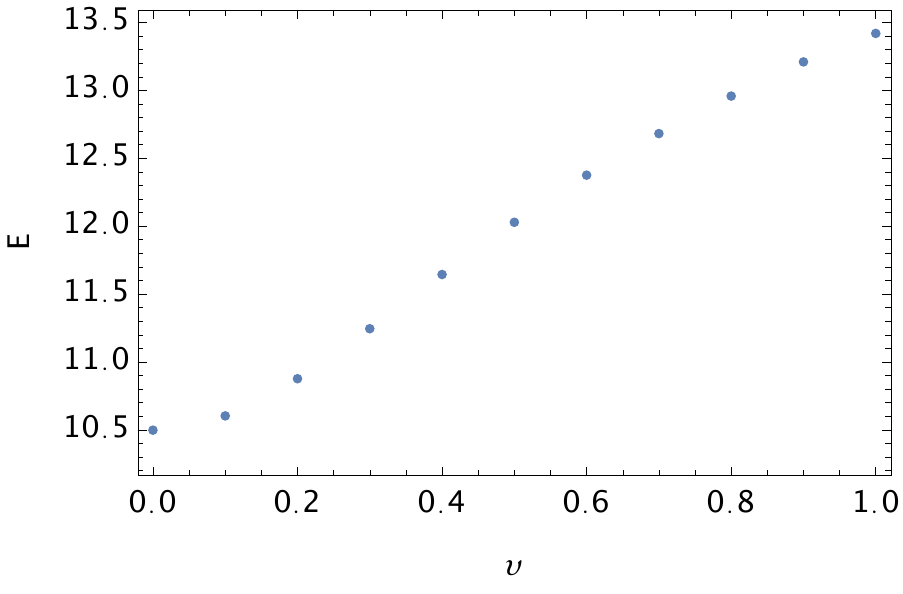} 
\caption{Plot of the energy as a function of $\nu$ for 4 particles. $\tilde\lambda$ is equal to 0.5. The simulation is performed at a temperature of $\tilde\beta=1$. We used $P=12$ beads and $2\times10^7$ MD steps to assure convergence. The statistical fluctuations are negligible for independent calculations. Hence, we did not give error bar in all our figures.}
\label{Anyon1}
\end{center}
\end{figure}

As a first test, we apply our method for general anyons by varying the parameter $\nu$ from 0 (bosons) to 1 (fermions), and plot the energy as a function of $\nu$, as shown in Fig. \ref{Anyon1} for $N=4$ particles. We used a nonzero interaction between particles to reduce sign problem \cite{HirshbergFermi,Dornheim}. The energies calculated are taken to be the real part of the complex average energy estimator presented above. The imaginary part of the energy estimator provides an indication for the accuracy of our simulations, a small imaginary part implies that our energy estimation is accurate. In all our simulations in this work, the imaginary part of the energy estimator is negligible. In our simulations, $|{Im E/Re E}|<0.002$. At $\nu=1$ for fermions, our result agrees with the numerical result in Ref. \cite{HirshbergFermi} based on PIMD. Hence, our simulations are correct in the whole region from bosons to fermions.
\par

\begin{figure}[htbp]
\begin{center}
 \includegraphics[width=0.75\textwidth]{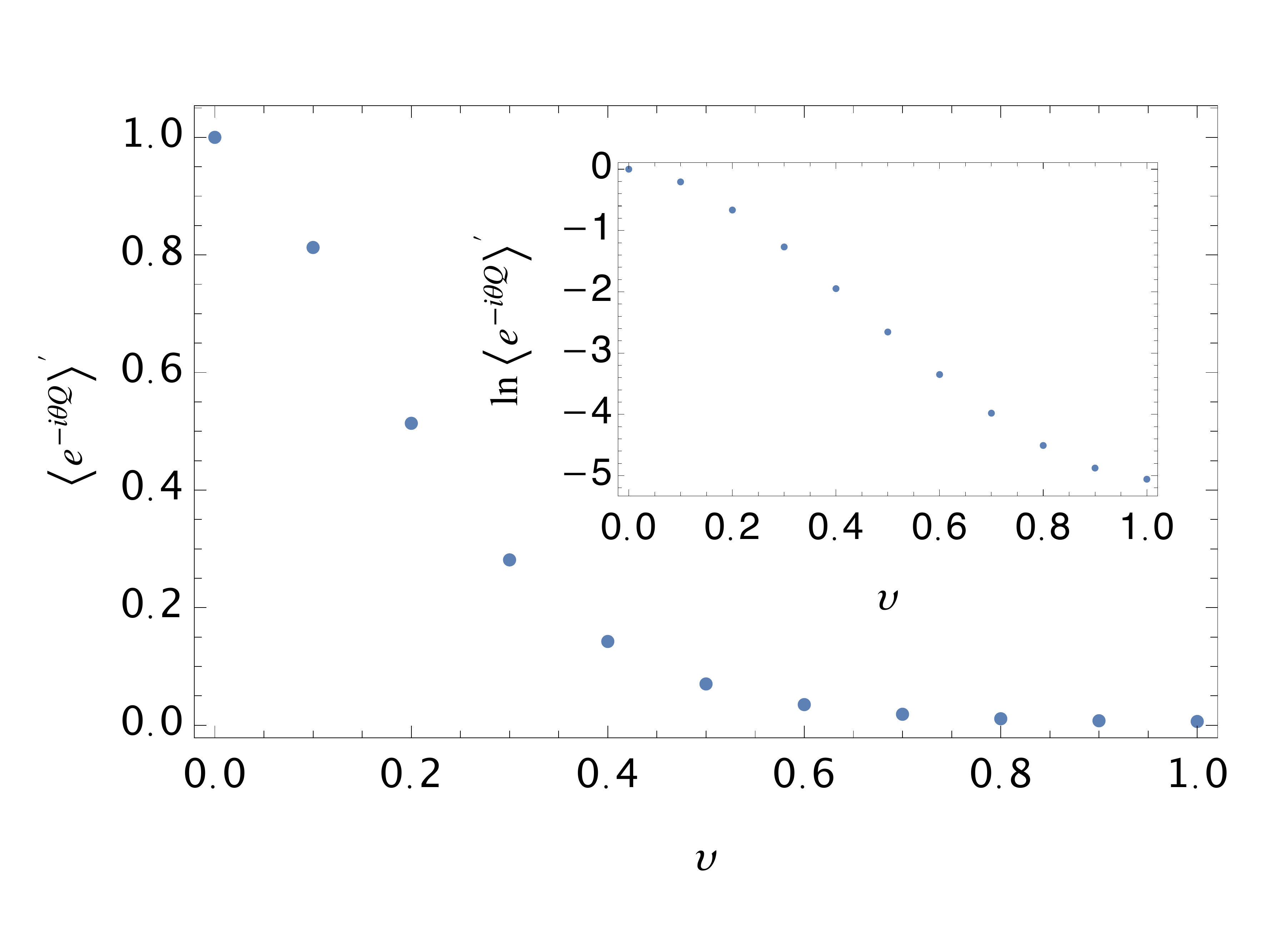} 
\caption{Plot of the magnitude of $\left<e^{-i\theta Q}\right>'$ as a function of $\nu$ for 4 particles. For bosons ($\nu=0$) $\left<e^{-i\theta Q}\right>'$ is always 1.0. As we increase $\nu$, the magnitude of $\left<e^{-i\theta Q}\right>'$ continuously drops to a small number, fermions ($\nu=1$) correspond to the smallest magnitude and most severe sign problem. The inset for $\ln\left<e^{-i\theta Q}\right>'$ shows that $\left<e^{-i\theta Q}\right>'$ decreases exponentially with increasing $v$.}
\label{Anyon2}
\end{center}
\end{figure}

In order to provide a further investigation into sign problem, we also plot the magnitude of $\left<e^{-i\theta Q}\right>'$ as a function of $\nu$, as shown in Fig. \ref{Anyon2}. A smaller magnitude means that the sign problem is more severe, and more MD iterations are needed to achieve convergence. We found that the sign problem is especially severe for larger number of particles and lower temperatures. To assure the accuracy, we used $2\times 10^7$ MD steps for importance sampling, and more MD steps were used to check the convergence.
\par

\begin{figure}[htbp]
\begin{center}
 \includegraphics[width=0.75\textwidth]{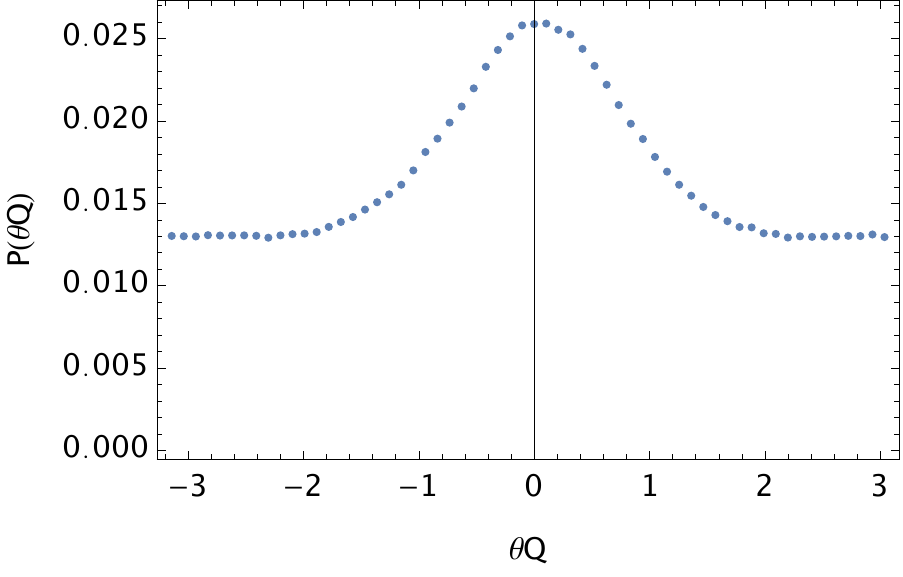} 
\caption{Distribution of the angle $\theta Q$, from $-\pi$ to $\pi$, for interacting fermions with $\tilde\lambda=0.5$.}
\label{Anyon3}
\end{center}
\end{figure}

We also plot the distribution of the angle $\theta Q$. For fermions with non-zero interaction, the distribution takes the form of a symmetric and Gaussian function, as illustrated in Fig. \ref{Anyon3}. In fact, the symmetric probability distribution of $\theta Q$ is the inevitable outcome of the path integral formalism of the partition function given by Eq. (\ref{Apartition}).
\par

\begin{figure}[htbp]
\begin{center}
 \includegraphics[width=0.75\textwidth]{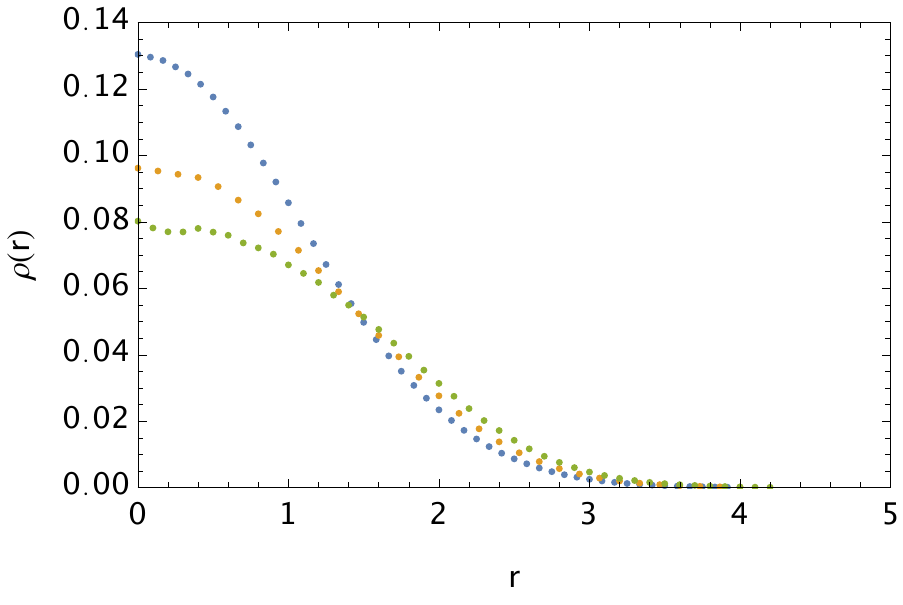} 
\caption{The density function for anyons, normalized to unity. Blue, orange and green represent $\nu=0$, $\nu=0.5$ and $\nu=1.0$ respectively. As $\nu$ increases the peak of the density decreases and the distribution of anyons becomes broader.}
\label{Anyon4}
\end{center}
\end{figure}

Finally, we also investigate the density function for anyons. As $\nu$ increases from 0 to 1, the energy increases, and therefore the density should become broader because of Pauli exclusion principle. This phenomenon is shown in Fig. \ref{Anyon4}.

\section{Calculation of the energy of fermions by anyon model}
\label{alleviate}

For anyon systems, it is clear that the sign problem becomes more severe with the increasing of $\nu$ between $0\leq \nu\leq 1$. In Fig. \ref{Anyon2}, it is shown clearly that $\left<e^{-i\theta Q}\right>'$ decreases exponentially with increasing $v$, which leads to the sign problem \cite{ceperley,loh,troyer,lyubartsev,vozn,Yao1,Yao2,Yao3}. In our numerical simulations, we can calculate more accurately the energy for $0\leq \nu\leq 0.5$, while less accurately for $\nu\geq 1/2$ (in particular $\nu=1$ for fermions) because of the sign problem. Another numerical inaccuracy in our simulations lies in the approximate calculation of $Q$. Nevertheless, it is expected that $E(\nu)$ is an analytical function of $\nu$ for finite number of particles. When numerical simulation is considered, the analyticity of $E(\nu)$ may require finite temperature so that $P$ is finite. In the thermodynamic limit or zero temperature with infinite $P$, because there are infinite number of beads, we can not assure the analyticity of $E(\nu)$. Of course, it is not necessary to worry about this in practical numerical simulation with finite number of beads. Hence, a reasonable fitting of $E(\nu)$ for small $\nu$ may give us accurate result for the energy of fermions $E(\nu=1)$.

From the partition function for anyons, it is quite interesting to notice that $E(\nu)$ is a periodic function \cite{Myrheim} of $\nu$ with period $2$. From the symmetry analysis, we also have $E(\nu)=E(-\nu)$. In addition, it is a monotonically increasing function of $\nu$ for $0\leq \nu\leq 1$ because of different exchange forces between bosons and fermions. The exchange forces for fermions are repulsive, while the exchange forces for bosons are attractive. This suggests the following simple function for $E(\nu)$.
\begin{equation}
E(\nu)=E(\nu=0)+a \left[\sin\left(\frac{\pi \nu}{2}\right)\right]^b.
\label{enu}
\end{equation}
Here $a$ and $b$ are two positive real numbers for fitting. The choice of $\sin\left(\frac{\pi \nu}{2}\right)$ in the above expression is due to the general consideration of $E(\nu)$.

\begin{figure}[htbp]
\begin{center}
 \includegraphics[width=0.75\textwidth]{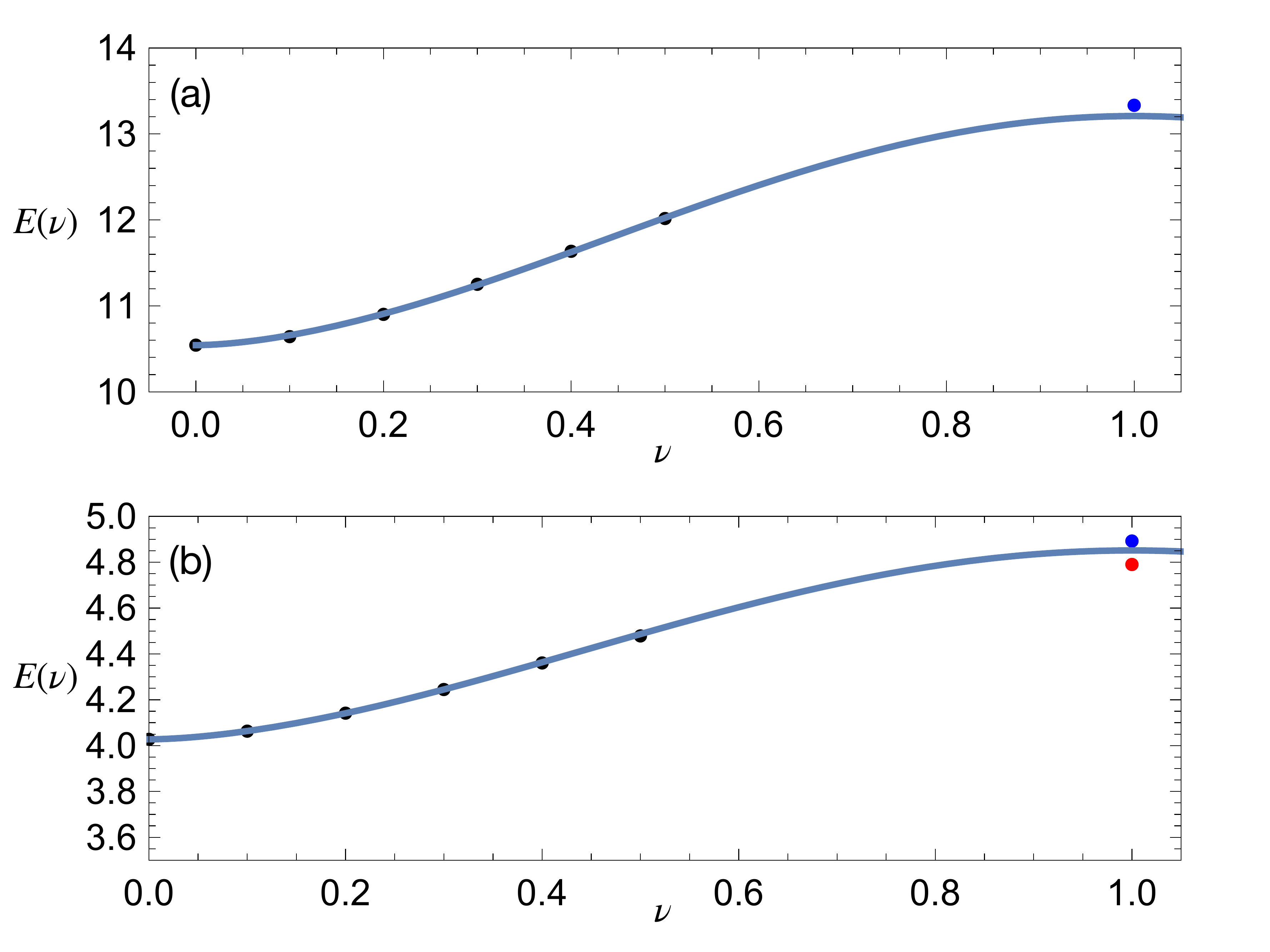} 
\caption{(a) For 4 interacting anyons with the same parameters of Fig. \ref{Anyon1}, the solid line illustrates the fitting with the data shown by black circles. The blue circle gives the numerical result of $E(\nu=1)$ for fermions, which agrees with the extrapolation in our fitting. (b) For two ideal anyons in a harmonic trap, the solid line shows the fitting of the data shown by black circles in our numerical simulation. As a comparison, the blue circle is the result of numerical simulation, while the red circle is the exact result. We see that the fitting to get the energy of fermions agrees better with the red circle. The slight difference is due to the numerical error in calculating $e^{-i\theta Q}$.}
\label{anyonsfitting}
\end{center}
\end{figure}

Because small value of $\nu$ has less severe sign problem, we may consider to calculate $E(\nu)$ below a value $\nu_c$, and use the above function to fit the data to get $E(\nu)$ for fermions with $\nu=1$.
For the same parameters in Fig.  \ref{Anyon1}, we use the numerical results of $E(0)$, $E(0.1)$, $E(0.2)$, $E(0.3)$, $E(0.4)$, $E(0.5)$ for data fitting. In Fig. \ref{anyonsfitting}(a), the solid line is the best fitting with function (\ref{enu}). We do find that the fitting can give accurate energy for fermions, while suffering from much less severe sign problem. The error of the fitting for the fermions is $2\%$, compared with the result of direct numerical simulation. If the whole region is considered, the mean deviation is smaller than $1\%$.

To test further our idea, we also consider here two ideal anyons which have analytical result for the partition function.
We consider here two anyons in a two-dimensional harmonic trap $\frac{1}{2}m\omega^2(x^2+y^2)$. In unit of $\hbar\omega$ and with the parameter $\tilde\beta=1$, in Fig. \ref{anyonsfitting}(b), the black circles give the average energy $E(\nu)$ as a function of $\nu$. As expected, we see a monotonic increase of the energy from $\nu=0$ for bosons to $\nu=1/2$ for anyons.
For this anyon model, we have analytical result of the energy spectrum. The analytical result of the partition function is \cite{Myrheim}
\begin{equation}
Z(\beta,\nu)=\frac{\cosh ((1-\nu)\beta)}{8\sinh^2(\frac{\beta}{2})\sinh^2 \beta}.
\end{equation}
The solid line in Fig. \ref{anyonsfitting}(b) gives the average energy based on the fitting with function (\ref{enu}). The red circle gives the energy of two fermions based on the analytical result of the partition function, which agrees well with the extrapolation of our fitting.  The error of the fitting for the fermions is $1\%$, compared the exact analytical result. If the whole region is considered, the mean deviation is smaller than $0.5\%$.
The slight difference is due to the systematic error of the numerical error of $Q$ because of the discretization in imaginary time and the function for fitting. In the limit of zero temperature, the derivative of $E(\nu)$ about $v$ is nonzero \cite{Myrheim} at $v=0$ and $v=1$, which is different from the situation of PIMD simulation with finite number of beads. Hence, we should consider carefully the validity of PIMD for zero temperature, when the practical application of our method is considered.

The above two examples show that we have hope to calculate the energy of fermions by anyon models with less severe sign problem. Of course, the present idea can not be applied to large fermion systems because in this case, there are still insurmountable obstacles due to the sign problem even for small $\nu$.

\section{summary and discussion}
\label{summary}
In this work we showed how to extend traditional PIMD simulations \cite{chandler,Parrinello,Tuckerman,Hirshberg,HirshbergFermi,Miura,Cao,Cao2,Jang2,Ram,Poly,Craig,Braa,Haber,Thomas} to enable the studies of general anyons. Through simple modifications of various physical estimators, we are able to extract physical quantities from simulations for a system of identical interacting anyons. Our methodology may be easily extended to study realistic physical systems of anyons with more complex interactions. We look forward to the applications of PIMD for anyons in the exciting field of many-body quantum systems.
\par

By developing a numerical algorithm for anyons, we also provide further insights into the fermion sign problem. Using the extra freedom in choosing the statistics parameter $\nu$, combined with the choice of interaction parameters, we may extrapolate regime in which the sign problem is less severe into regime with most severe sign problem \cite{HirshbergFermi,DornheimMod} (i.e. weakly interacting fermions), hopefully attenuating the fermion sign problem further. Although the present method based on anyon models can not be applied to large fermion system, our work suggests that we may consider to construct more appropriate fictitious particle \cite{XiongSFP} with a parameter to connect bosons and fermions, similarly to anyon models connecting continuously bosons and fermions.

\textbf{Acknowledgements} This work
is partly supported by the National Natural Science Foundation of China under grants number 11175246, and 11334001.

\textbf{DATA AVAILABILITY}

The data that support the findings of this study are available from the corresponding author upon reasonable request.
The code of this study is openly available in GitHub (https://github.com/xiongyunuo/PIMD-Anyon).



\end{document}